\documentclass{article}
\usepackage{spconf,amsmath,graphicx,hyperref}
\usepackage{cancel}

\usepackage{color}
\usepackage[dvipsnames]{xcolor}

\title{Evaluating High-Resolution Piano Sustain Pedal Depth Estimation with Musically Informed Metrics}
\name{Hanwen Zhang$^{1*}$, Kun Fang$^{1*}$, Ziyu Wang$^{2,3}$, Ichiro Fujinaga$^{1}$\thanks{$^*$Equal contribution. Both authors were supported in part by funding from the Social Sciences and Humanities Research Council of Canada (SSHRC) under Grant No. 895-2022-1004. Code: \url{https://github.com/kunfang98927/PedalDetection/blob/icassp2026/}}}
\address{%
$^{1}$ Schulich School of Music, McGill University, Montréal, QC, Canada\\
$^{2}$ Courant Institute of Mathematical Sciences, New York University, New York, USA\\
$^{3}$ Mohamed bin Zayed University of Artificial Intelligence (MBZUAI), Abu Dhabi, UAE
}

\begin{document}
\ninept
\maketitle
\begin{abstract}
Evaluation of continuous piano pedal depth estimation tasks remains incomplete when relying only on conventional frame-level metrics, which overlook musically important features such as direction-change boundaries and pedal curve contours. To provide more interpretable and musically meaningful insights, we propose an evaluation framework that augments standard frame-level metrics with an \textit{action}-level assessment measuring direction and timing using segments of \emph{press}/\emph{hold}/\emph{release} states and a \textit{gesture}-level analysis that evaluates contour similarity of each pedal press-release cycle. We apply this framework to compare an audio-only baseline with two variants: one incorporating symbolic information from MIDI, and another trained on binary-valued targets (pedal on/off only), all within a unified architecture. Results show that the MIDI-informed model significantly outperforms the others at action and gesture levels, despite modest frame-level gains. These findings demonstrate that our framework captures musically relevant improvements indiscernible by traditional metrics, offering a more practical and effective approach to evaluating pedal depth estimation models.
\end{abstract}
\section{Introduction}
\label{sec:intro}

Although the sustain pedal is a continuous control crucial for producing distinct sound effects essential to classical piano playing, previous studies treated it as an auxiliary signal within piano transcription systems \cite{Hawthorne2018, kong_high-resolution_2021, yan_skipping_2021, yan_scoring_2024} and mostly focused on binary on/off detection \cite{liang_detection_2017, liang_piano_2017, liang_measurement_2018, liang_piano_2018, liang_piano_2019, liang_transfer_2019}.
Previous work \cite{Fang25pedal} demonstrates the importance and benefits of continuous pedal depth estimation. However, compared with mature MIR tasks such as melody extraction, beat tracking, and chord recognition, pedal depth estimation remains underexplored, and its evaluation still largely relies on standard generic frame-level metrics (MSE, MAE, F1, etc.), which are not specifically designed for musically meaningful information, such as whether action boundaries are correct and whether phrase-level contours are reasonable. Binary F1-score can ignore important cases such as half-pedal and pedal fluttering, while multi-class F1-score overlooks expressive musical aspects of the pedal curve. Frame-level regression metrics (MSE/MAE) better reflect continuous values, but still over-penalize small timing shifts or minor jitters, even when two pedal curves are nearly equivalent in perceived musical effect.

\begin{figure}[t!]
\centering
\includegraphics[width=0.5\textwidth]{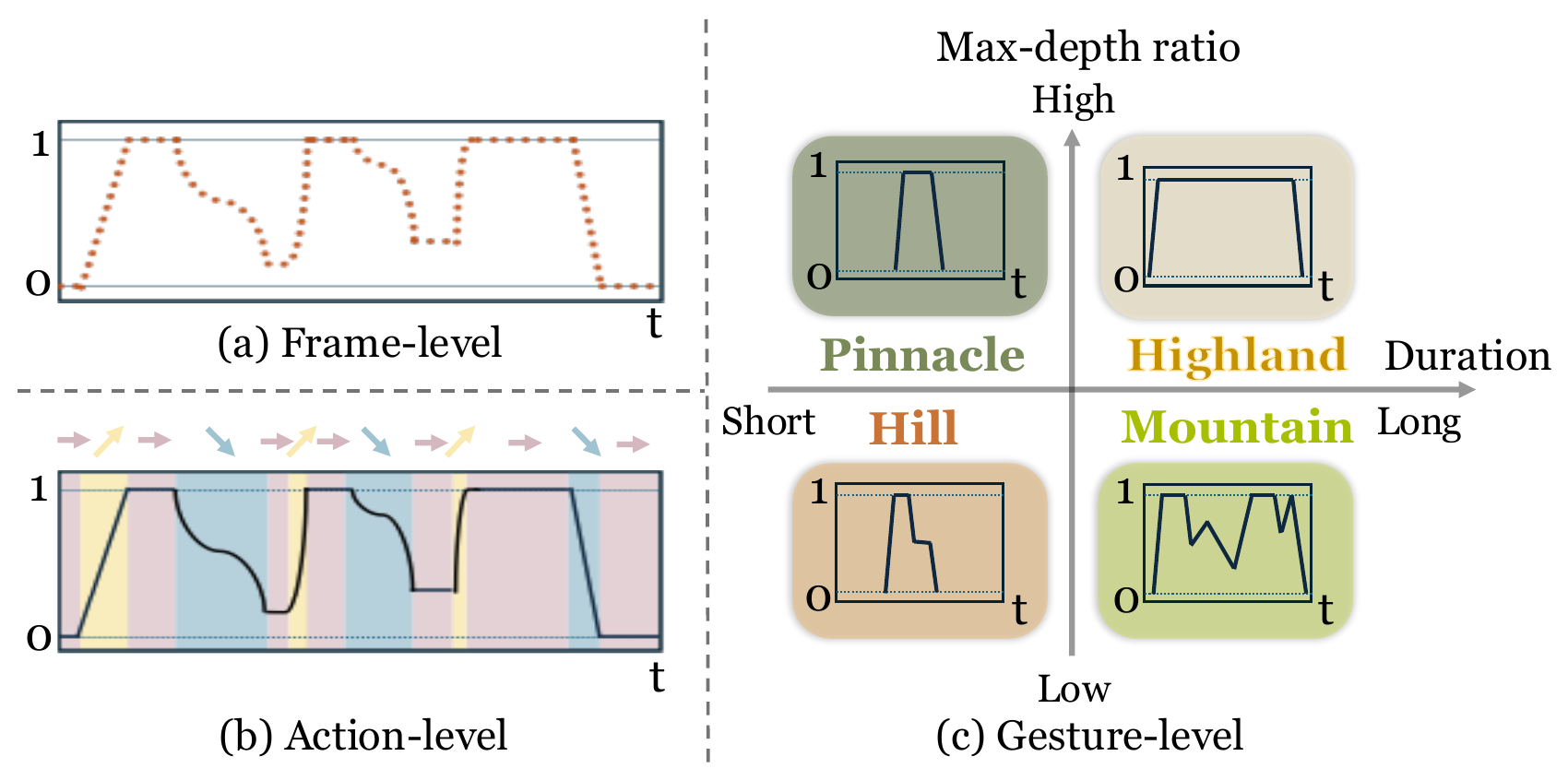}
\caption{An illustration of the three-level analysis proposed and used in this work. (a) Frame: each time step is treated as the basic unit. 
(b) Action: consecutive frames sharing the same direction/state are merged into segments labeled \emph{press} (yellow), \emph{hold} (pink), or \emph{release} (blue). 
(c) Gesture: complete pedal press-release cycles, characterized and classified according to canonical contour shapes.}
\label{fig:figure1}
\end{figure}

In this paper, we propose evaluating pedal depth estimation models not only with existing frame-level metrics but also using two additional levels of metrics that are musically more interpretable and practically more informative: (1) an \textit{action-level} evaluation, which assesses predictions based on \emph{press}/\emph{hold}/\emph{release} segments; and (2) a \textit{gesture-level} evaluation, which treats each press–release cycle as a unit and compares contour similarity within that unit. Evaluation at action and gesture levels extends the generic frame-based analysis by providing richer insights into model performance that align with actual practice and pedagogical concepts in piano playing.

To validate our proposed evaluation framework, we compare three model variants with expected capability gaps: audio-only (baseline), MIDI-informed, and audio-only but trained on binarized (on/off) targets, all built on the same Transformer-based architecture. The goal is to test whether our framework can reveal performance differences that frame-level metrics fail to capture, and whether action- and gesture-level evaluation can better characterize model behavior in musical contexts.
Results show that the binary-target model underperforms at predicting action states and capturing gesture shapes;
adding MIDI to baseline further strengthens action boundary clarity and gesture contour matching, though fine micro-dynamics in long \emph{mountain}-like gestures remain challenging for our models. Overall, this framework provides more musically meaningful analysis and thus more informative diagnosis of model performance for piano sustain pedal depth estimation tasks.

\section{Related Work}
\label{sec:format}

Early sustain pedal depth estimation research projects evolved from sensor-based, small-scale classification \cite{liang_detection_2017,liang_piano_2017,liang_piano_2018,liang_measurement_2018} to binary pedal detection within transcription systems \cite{liang_piano_2019,liang_transfer_2019,kong_high-resolution_2021,yan_skipping_2021,yan_scoring_2024}, but evaluation still relies on frame-level metrics. \cite{Fang25pedal} demonstrated feasibility but further underscored the absence of a unified, reproducible evaluation paradigm, prompting us to reconsider both what to evaluate and how, via a musically informed approach.

More broadly, temporal music information retrieval (MIR) research has developed a family of task-aligned evaluation protocols across continuous and discrete, low- and high-level targets \cite{raffel2014_mir_eval}. Continuous signals (e.g., melody) are typically scored by frame-wise accuracy \cite{poliner2007_melody_transcription, salamon2014_melody_extraction}. Discrete events (e.g., onsets, beats) use tolerance-window precision/recall/F1 and continuity measures \cite{davies2009_beat_eval_tr, bock2012_onset_online}. Chord recognition, structural segmentation, and pattern discovery are segment/structural tasks; accordingly, evaluation typically emphasizes duration-weighted overlap, vocabulary mappings, and pairwise consistency, and specifically for segmentation adopts a hierarchical scheme that separates boundary localization from structural labeling \cite{pauwels2013_evaluating_chords, ni2013_subjectivity_chord_eval, turnbull2007_supervised_boundaries, levy2008_struct_seg_clustering, lukashevich2008_segmentation_measures}. Following this trajectory, we position pedal depth estimation as a composite temporal task that simultaneously involves a continuous curve (depth), discrete events (press/release boundaries), and structural semantics (contours). It therefore requires measurements at the corresponding levels and a unified aggregation, rather than a single-scale, generic metric.

\section{Pedal Prediction Baselines}
\label{sec:pagestyle}

Before introducing our proposed pedal evaluation methods, we first describe the models used for comparison and present conventional metrics. Section~\ref{subsec:model} defines the task and introduces controlled model variants based on Transformer; 
Section~\ref{subsec:frameeval} reviews standard frame-level metrics; 
Section~\ref{subsec:challenge} discusses their limitations, thus indicating the need for our music-informed evaluations in Section~\ref{subsec:framework}.

\subsection{Pedal Prediction Model Variants}
\label{subsec:model}

We formulate sustain pedal prediction as a temporal sequence estimation task. We consider two input representations: \emph{audio} denotes log-mel spectrograms (229 bins) and MFCCs (20 dims) computed on $\sim$5\,s windows (500 frames), while \emph{MIDI} denotes frame-aligned 88-dim note-velocity vectors obtained from audio via transcription using \cite{yan_scoring_2024}. The model outputs a frame-wise continuous pedal depth sequence $x_{1:T}\!\in\![0,1]$ (MIDI CC64 normalized by 127), frame-wise pedal onset $o_{1:T}$ and offset $f_{1:T}$ binary event sequences, and a segment-level global depth $g\!\in\![0,1]$ that equals the segment pedal level average using the following loss:
\begin{equation}
\mathcal{L}_{\text{total}}
= \lambda_1 \mathcal{L}_{\text{pedal}}
+ \lambda_2 \mathcal{L}_{\text{global}}
+ \lambda_3 \mathcal{L}_{\text{onset}}
+ \lambda_4 \mathcal{L}_{\text{offset}},
\end{equation}
where $\mathcal{L}_{\text{pedal}}$ and $\mathcal{L}_{\text{global}}$ are MSE losses on frame-wise and segment-level depth, and $\mathcal{L}_{\text{onset}}$, $\mathcal{L}_{\text{offset}}$ are BCE losses on event sequences; the weights $\lambda_{1..4}$ are fixed across variants.

Building on \cite{Fang25pedal}, we keep multi-task heads and streamline the input stage: mel features are encoded by a small CNN, MFCCs by an MLP, the fused representation is passed to a Transformer encoder (8 heads, standard FFN), and the output heads produce the targets above. To isolate the roles of output format and input modality, we train three controlled variants under identical architecture, hyperparameters, and optimization:
\begin{enumerate}
    \item \textsc{Audio (Binary)}:  binarized labels trained with BCE; raw sigmoid output (before thresholding) used as predicted depth.
    \item \textsc{Audio}: baseline continuous regression on audio features (mel bins {+} MFCCs).
    \item \textsc{Audio+MIDI}: same as \textsc{Audio}, with an additional fused, frame-aligned MIDI (pitch and velocity) stream.
\end{enumerate}






\subsection{Frame-Level Evaluation}
\label{subsec:frameeval}

Standardized evaluation protocols for continuous pedal estimation are still under discussion; most evaluations still borrow criteria from binary prediction \cite{Fang25pedal}. Conventional frame-wise metrics measure: (i) classification scores after discretization (binary and 4-class) using Precision/Recall/F1, and (ii) regression errors on the continuous depth $x_{1:T}$ using MAE and MSE. 
These summarize per-frame signal fidelity and enable comparison to prior work.

\subsection{Challenges in Existing Evaluation}
\label{subsec:challenge}

We highlight the following issues and limitations of frame-based metrics for pedaling: (i) Not all frames matter equally: boundary frames at harmonic changes or intended releases are musically critical, whereas interior frames within a sustained harmony can tolerate substantial flutter. Uniform frame weighting over-penalizes acceptable fluctuations and under-emphasizes boundary accuracy. (ii) Contour insensitivity: frame-wise metrics measure absolute differences but ignore the general contour: phase-shifted yet musically equivalent curves and equal long holds with different micro-oscillations can receive large errors despite similar intent. (iii) Low diagnostic value: aggregate frame scores do not indicate what to improve (timing, duration, contour) or where failures occur (which gesture types), offering limited guidance for model design and error analysis.

\begin{figure}[h!]
\centering
\includegraphics[width=0.5\textwidth]{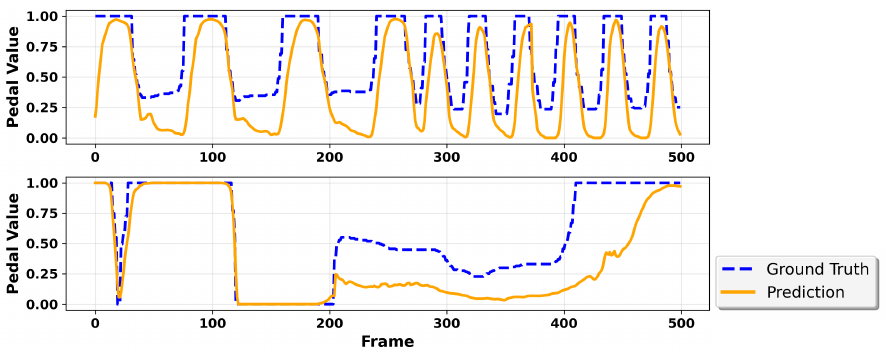}
\caption{Two examples with large MSE and MAE but correctly following the pattern. The top one has MSE 0.0983 and MAE 0.2425. The bottom one has MSE 0.0866 and MAE 0.2053.}
\label{fig:bad example}
\end{figure}

These observations motivate the music-informed analysis introduced next (Sections~\ref{subsec:action} and \ref{subsec:gesture}), which emphasizes boundary correctness, segment coverage, and contour similarity. Figure \ref{fig:bad example} showcases two excerpts where the MSE and MAE point at relatively poor predictions, even though the general pattern is captured and aligned properly in time.

\vspace{-5pt}
\section{Evaluation Framework}
\label{subsec:framework}

Beyond frame-level metrics, we propose two novel perspectives for evaluation inspired by piano pedagogy: (i) an action-based evaluation (Section~\ref{subsec:action}), in which we segment the pedal curve into \textit{press}/\textit{hold}/\textit{release} and score temporal coverage and alignment; and (ii) a gesture-based evaluation (Section~\ref{subsec:gesture}), where we consider complete pedal gestures from pedal on to off and compare characteristic contour shapes that imply pedaling in expressive piano performance.

\subsection{Action-Based Evaluation}
\label{subsec:action}


Instead of working with quantized or raw depth values, we now treat pedaling as a sequence of discrete directional actions (\emph{press}, \emph{hold}, and \emph{release}) derived from the sustain pedal signal $x_{1:T}\in[0, 1]$. This representation reflects performers’ intentional control of the pedal at different musical moments. As noted in lessons with Claudio Arrau \cite{vonArx2014}, pianists and teachers pay close attention to pedaling and predominantly describe it in terms of actions. Compared to exact depth values, the exact timing and type of action better correspond to how pianists conceptualize the use of sustain pedal.


However, ground truth signals from optical sensors and model outputs inevitably contain local jitter. To preserve original temporal resolution while avoiding purely frame-level analysis, we run linear regression on a sliding window centered on each frame. We use a window size of $19$, a slope threshold of $0.005$, and a minimum $R^2$ of 0.5 to identify action states for both ground truth and model predictions. Therefore, for each frame at time $t$, its action state is determined by the regression slope of the segment centered at $t$.
The action-based evaluation can now be framed as a three-class classification task for \textit{press}, \textit{hold}, \textit{release} classes using action states extracted using the steps mentioned above,
and the evaluation follows the standard precision, recall, and F1 framework. 
We report scores for each action class as well as class-balanced macro and weighted averages, all shown in Table~\ref{tab:action}.

\subsection{Gesture-Based Evaluation}
\label{subsec:gesture}





Prior work \cite{Fang25pedal} shows sustain pedaling is not a binary down–up switch but nuanced press–to–release cycles whose shapes depend on musical intent. Therefore, we propose \emph{gestures} as an important perspective to evaluate pedal depth estimation. We therefore evaluate from the perspective of \emph{gestures}, defined as contiguous segments that begin when depth first exceeds a small threshold $\epsilon$ (press onset) and end at the subsequent return below $\epsilon$ (release). Formally, let $x_{1:T} \in [0, 1]$ be the frame-wise pedal depth at a fixed frame rate. A gesture $g=x_{t:t+n}$ is a maximal interval satisfying:
\begin{equation*}
    x_{t-1}\le \epsilon\text{, } x_{t+n+1} \le \epsilon\text{, and } x_j > \epsilon \, \forall j\in[t, t+n]\text{.}
\end{equation*}
That is, $g$ starts at the first frame exceeding $\epsilon$, remains above $\epsilon$, and ends at the first return below $\epsilon$. To classify various types of gestures, we further compute a \textit{max-depth ratio}, which measures how long the pedal stays or fluctuates near its maximum depth within the segment:
\[
r(g) = \frac{\bigl|\{x_j \ge \theta\max(g) |x_j \in g\}\bigr|}{n}\text{.}
\]

For a given piece, the pedal value sequence can thus be segmented into gestures, which are then classified along two axes, \emph{duration} and \emph{max-depth ratio} $r(g)$, into four canonical shapes (illustrated in Figure~\ref{fig:figure1}).
\begin{itemize}
  \item \textbf{\textit{Pinnacle}}: (short, high $r$): a quick press–release with a dominant peak; used for brief accents or as catch pedal.
  \item \textbf{\textit{Hill}}: (short, low $r$): a short press–release with gentle shaping (e.g., partial releases/half-pedal control); more decorative and nuanced than \textit{Pinnacle}.
  \item \textbf{\textit{Highland}}: (long, high $r$): press–hold–release with an extended plateau, building resonance and/or sound volume (common in large Romantic piano repertoires).
  \item \textbf{\textit{Mountain}}: (long, low $r$): sustained depth with modulation/oscillation before release, enabling extended blending and color (typical in Impressionist compositions).
\end{itemize}
The remaining intervals (i.e., those with depth $< \epsilon$ in between the four gestures listed above) are labeled as \textit{plain}, a category that does not correspond to any gesture type.


Classifying pedal gestures and evaluating them separately provides a clearer and more musically meaningful interpretation. Pedaling, as Banowetz describes, is a “downward journey” \cite{banowetz_pianists_1985}, and contiguous pedal cycles, which we define as \textit{gestures}, appear to be the most appropriate unit for evaluating pedal depth estimation in a musical context.

For all  non-\emph{plain} gestures in the corpus:
duration (frames) has mean $147.8$, median $70.0$, std $402.9$; max-depth ratio has mean $0.645$, median $0.688$, std $0.220$. We set max-depth ratio threshold to $0.65$ and duration threshold to $100$ frames to define our quadrants.

We already see that frame-wise scores can miss contour similarity in Section \ref{subsec:challenge}. To compare gesture shapes while tolerating small local deviations, we adopt (i) Fourier-descriptor analysis: compute discrete Fourier coefficients and reconstruct with the first $K{=}11$ coefficients to suppress high-frequency noise/model artifacts, and compare the reconstructed low-pass signals via MSE; (ii) 5-point analysis: compute duration-weighted MSE using five landmarks per gesture $x_{1:T}$: $x_1$, $x_T$, $\text{median}(x_{1:T})$, $\text{mean}(x_{1:T})$, $\text{max}(x_{1:T})$.
According to these new metrics, the top example in Figure \ref{fig:bad example} yields an MSE of 0.0586 using the Fourier method, whereas the bottom example obtains an MSE of 0.0274 in the 5-point analysis, now both being less harshly graded than by raw MSE scores.

\section{Experiments}
\label{sec:typestyle}

This section presents our experiments with three models detailed in \ref{subsec:model} and their performance evaluation using our proposed new framework: Section~\ref{subsec:dataset} details the dataset and unified training setup; Section~\ref{subsec:across-models} shows results and findings across models; Section~\ref{subsec:strengths-limitations} summarizes consistent strengths and remaining limitations.

\subsection{Dataset and Training}
\label{subsec:dataset}

We use MAESTRO v3.0.0 \cite{hawthorne_enabling_2019}, which, among other piano corpora \cite{asap-dataset,emiya:inria-00544155}, is well suited for sustain pedal depth estimation for its professional recordings, expert performances, a curated repertoire, and synchronized optical pedal depth.

All models are trained on a single NVIDIA H100 (80\,GB) with batch size 32, using AdamW 
(\(\beta_1{=}0.9,\ \beta_2{=}0.999\), weight decay \(0.01\)) and a OneCycle scheduler 
(peak LR \(5{\times}10^{-4}\); 10\% warm-up; init factor \(1/25\); final factor \(1/100\); cosine). We train for 15 epochs and select the best checkpoint (\(\sim\)150k steps, epoch 13). To ensure fair comparison, all hyperparameters and architecture are shared across variants, except the additional MIDI input.

\begin{table}[ht]
\small
\centering
\footnotesize
\setlength{\tabcolsep}{1.4pt} 
\renewcommand{\arraystretch}{1.1} 
\begin{tabular}{lccc|ccc|cc}
\hline
Model & \multicolumn{3}{c|}{Binary (P$\uparrow$/R$\uparrow$/F1$\uparrow$)} & \multicolumn{3}{c|}{4-Class (P$\uparrow$/R$\uparrow$/F1$\uparrow$)} & MSE $\downarrow$ & MAE $\downarrow$\\
\hline
\cite{Fang25pedal} & 0.8975 & 0.8971 & 0.8973 & 0.6849  & 0.6971  & 0.6863  & 0.0425 & 0.1339\\
\textsc{Audio (Binary)} & 0.8946 & 0.8944 & 0.8945 & 0.6016  & 0.6627  & 0.6166  & 0.0582 & 0.1502 \\
\textsc{Audio} & 0.9043 & 0.9037 & 0.9039 & 0.7013  & 0.7153  & 0.7045  & 0.0416 & 0.1237 \\
\textsc{Audio+MIDI} & 0.9379 & 0.9370 & 0.9372 & 0.7529  & 0.7661  & 0.7546  & 0.0280 & 0.0986 \\
\hline
\end{tabular}
\caption{Frame-level results.}
\label{tab:quantitative_results}
\end{table}

\begin{table*}[t]
\centering
\small
\footnotesize
\setlength{\tabcolsep}{3pt} 
\renewcommand{\arraystretch}{0.96} 
\setlength{\tabcolsep}{4pt}
\begin{tabular}{lccc|ccc|ccc|cc}
\hline
& \multicolumn{3}{c}{\textit{Press}} & \multicolumn{3}{c}{\textit{Hold}} & \multicolumn{3}{c}{\textit{Release}} & \multicolumn{2}{c}{\textit{Overall}}  \\
Model & P $\uparrow$ & R $\uparrow$ & F1 $\uparrow$ & P $\uparrow$ & R $\uparrow$ & F1 $\uparrow$ & P $\uparrow$ & R $\uparrow$ & F1 $\uparrow$ & F1 (Macro) $\uparrow$ & F1 (Weighted) $\uparrow$\\
\hline
\textsc{Audio (Binary)} & 
0.5068 & 0.6605 & 0.5739 &
0.8734 & 0.7961 & 0.8330 &
0.5380 & 0.6347 & 0.5823 &
0.6629 & 0.7655 \\
\textsc{Audio} & 0.5383 & 0.6959 & 0.6070 & 0.8877 & 0.8028 & 0.8431 & 0.5544 & 0.6850 & 0.6128 & 0.6876 & 0.7815\\
\textsc{Audio+MIDI} & 0.6325 & 0.7745 & 0.6964 & 0.9171 & 0.8568 & 0.8859 & 0.6807 & 0.7721 & 0.7235 & 0.7686 & 0.8392\\
\hline
\end{tabular}
\caption{Action-level results: per-action Precision/Recall/F1, plus macro- and weighted-F1 overall.}
\label{tab:action}
\end{table*}

\begin{table*}[t]
\centering
\small
\footnotesize
\setlength{\tabcolsep}{3pt} 
\renewcommand{\arraystretch}{1} 
\setlength{\tabcolsep}{3pt}
\begin{tabular}{l
cc|cc|cc|cc|cc|cc}
\hline
& \multicolumn{2}{c}{\textit{Mountain}} & \multicolumn{2}{c}{\textit{Highland}} & \multicolumn{2}{c}{\textit{Hill}} & \multicolumn{2}{c}{\textit{Pinnacle}} & \multicolumn{2}{c}{\textit{Plain}} & \multicolumn{2}{c}{Weighted} \\
Model & 5-pts $\downarrow$ & Fourier $\downarrow$
      & 5-pts $\downarrow$ & Fourier $\downarrow$
      & 5-pts $\downarrow$ & Fourier $\downarrow$
      & 5-pts $\downarrow$ & Fourier $\downarrow$
      & 5-pts $\downarrow$ & Fourier $\downarrow$
      & 5-pts $\downarrow$ & Fourier $\downarrow$ \\
\hline
\textsc{Audio (Binary)}     
 & 0.0863 & 0.0544 & 0.1010 & 0.0207 & 0.1163 & 0.0761 & 0.1136 & 0.0657 & 
0.1395 & 0.0512 & 0.1085 & 0.0460 \\
\textsc{Audio}      & 0.0689 & 0.0284 & 0.0880 & 0.0146 & 0.0899 & 0.0521 & 0.0941 & 0.0503 & 0.1332 & 0.0471 & 0.0946 & 0.0329 \\
\textsc{Audio+MIDI} & 0.0457 & 0.0273 & 0.0441 & 0.0116 & 0.0511 & 0.0358 & 0.0518 & 0.0291 & 0.0737 & 0.0247 & 0.0530 & 0.0225 \\
\hline
\end{tabular}
\caption{Gesture-level results: MSE for each gesture category and overall, using two shape measures: (i) 5-point, and (ii) Fourier (11 coefficients).}
\label{tab:gesture}
\end{table*}

\begin{figure}[t]
\centering
\includegraphics[width=1.\linewidth]{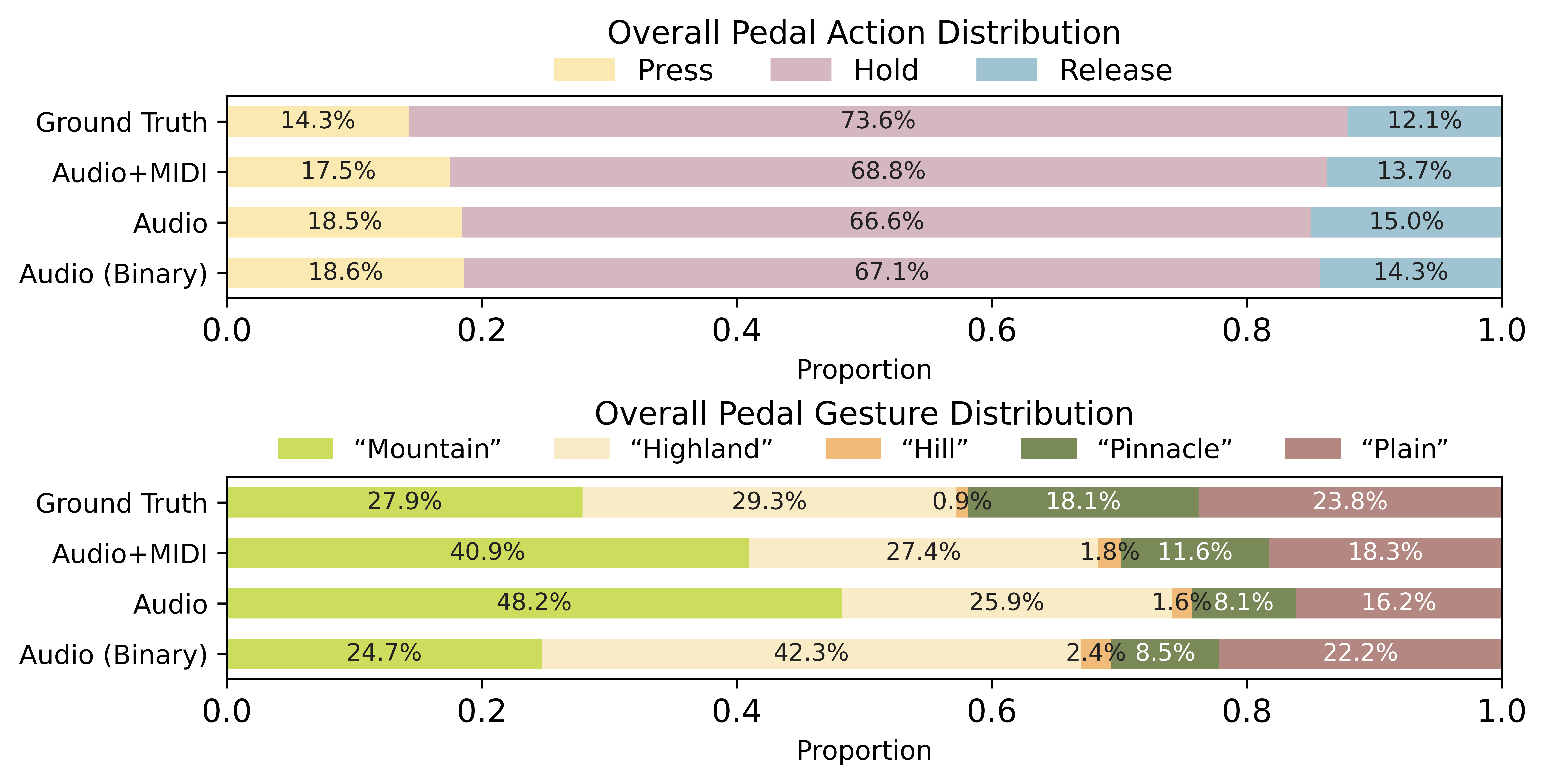}
\caption{Overall distribution of pedal \emph{actions} (top) and \emph{gestures} (bottom) across models. Each row corresponds to one model and the ground truth; bars are stacked by category, with percentages computed from summed frame counts in each category. Labels show per-category proportions.}
\label{fig:bar_overview}
\end{figure}

\subsection{Evaluation Across Models}
\label{subsec:across-models}

We use conventional frame-based metrics (Table~\ref{tab:quantitative_results}) as well as our action- (Table~\ref{tab:action}) and gesture-based metrics (Table~\ref{tab:gesture}). Overall, the \textsc{Audio+MIDI} model performs the best, followed by \textsc{Audio}, with \textsc{Audio (Binary)} last. This pattern highlights that estimating continuous pedal values is beneficial, and adding MIDI input further enhances performance by providing helpful structural priors.

\textbf{Continuous estimation matters (vs. binary classification).}
Although \textsc{Audio (Binary)} achieves a comparable frame-wise F1, it exhibits systematic bias at both action and gesture levels (Figure~\ref{fig:bar_overview}, Table~\ref{tab:action}).
Despite nearly identical action distributions, \textsc{Audio (Binary)} achieves lower F1 scores especially for \textit{press} and \textit{release} actions. This gap likely arises because continuous estimation captures the smooth depth changes during transitions, whereas binary classification reduces them to abrupt on/off switches.
At gesture level, \textsc{Audio (Binary)} largely collapses to predicting \textit{highland}, and rarely recognizes more complex patterns such as \textit{mountain} and \textit{pinnacle} (Fig.~\ref{fig:bar_overview}). Even for \textit{highland}, which \textsc{Audio (Binary)} predicts frequently, gesture-level MSE remains unsatisfactory (Table~\ref{tab:gesture}).

Models trained for continuous estimation have remarkably improved action-level recalls and modestly but consistently higher precisions than \textsc{Audio (Binary)} (Table~\ref{tab:action}). In addition, at the gesture level, \textsc{Audio} and \textsc{Audio+MIDI} can identify gesture categories other than \textit{highland} and \textit{pinnacle}, yielding distributions closer to ground truth (Figure~\ref{fig:bar_overview}), and they also result in substantially lower per-gesture MSE, a positive sign for more accurate shape modeling. These findings agree with \cite{Fang25pedal} and reinforce the importance of continuous-valued prediction for piano pedal estimation.


\textbf{Adding MIDI input provides structural priors.}
\textsc{Audio+MIDI}, on the other hand, predicts all gesture categories with a distribution the closest to ground truth (Figure~\ref{fig:bar_overview}). It also preserves shape more faithfully within each gesture. In particular, gesture-level MSE drops notably for short, rapid gestures such as \textit{pinnacle} and \textit{hill}, indicating greater temporal sensitivity than the \textsc{Audio} baseline (Table~\ref{tab:gesture}). The substantially improved \emph{5-point} scores and the modestly better \emph{Fourier} (11-coefficient) scores compared to \textsc{Audio} suggest that the extra MIDI input mainly helps predict the general trend more than high-frequency micro-oscillations. In addition, scores across gestures are more balanced for \textsc{Audio+MIDI}, which suggests that MIDI input helps the model generalize more consistently for both short and long gestures.

\label{subsec:strengths-limitations}

\subsection{Strengths and Limitations of Current Pedal Models}
\label{subsec:strengths-limitations}
We summarize consistent patterns across models: where they succeed reliably and where errors persist.

\textbf{Strengths.} \textit{Highland} is consistently the best-predicted gesture: its distribution rate is the closest to ground truth (Figure~\ref{fig:bar_overview}), and among all gestures, segments identified as \textit{highland} almost always have the lowest MSE according to Fourier and 5-point analyses (Table~\ref{tab:gesture}). This indicates that models are effective at capturing gestures with relatively stable, long pedal movements, where the resulting sound is more pronounced.

\textbf{Limitations and biases.} (i) Under-detection of high max-depth gestures. Models tend to under-predict \emph{Pinnacle}/\emph{Highland} (missed or confused with \emph{Hill}/\emph{Mountain}), suggesting difficulty with gestures having high max-depth ratios.
(ii) Short-and-fast gestures remain hard. \emph{Pinnacle}/\emph{Hill} are better captured but still with relatively high MSE, likely because the brief duration and rapid changes exceed the temporal resolution that per-frame prediction can reliably resolve. 


\section{Conclusion}
We studied sustain pedal depth estimation with controlled experiments and evaluated using our proposed new framework with three levels at \emph{frame}, \emph{action}, and \emph{gesture}. Across three model variants, we demonstrated that continuous-valued estimation is crucial for accurately capturing actions and preserving gesture characteristics. Pre-extracted MIDI pitch and velocity information provides useful structural priors for better contour shaping in gestures. While conventional frame-wise metrics can miss musically acceptable behavior, our proposed three-level approach provides a more comprehensive view of strengths and limitations of our models. These results show the value of our musically informed evaluation and point to future work on boundary-sensitive objectives, contour-aware losses, and perceptual validation for piano sustain pedal depth estimation tasks.


\vfill\pagebreak

\bibliographystyle{IEEEbib}
\bibliography{strings,refs}

@inproceedings{hawthorne_enabling_2019,
	address = {New Orleans, Louisiana, United States},
	title = {Enabling Factorized Piano Music Modeling and Generation with the {MAESTRO} Dataset},
	eventtitle = {{ICLR}},
	booktitle = {Proceedings of the 7th International Conference on Learning Representations},
	author = {C. Hawthorne and A. Stasyuk and A. Roberts and I. Simon and C.-Z. A. Huang and S. Dieleman and E. Elsen and J. Engel and D. Eck},
	date = {2019-01-17},
	year = {2019},
        address = {New Orleans, Louisiana, United States}
}

@inproceedings{liang_piano_2019,
	location = {Brighton, United Kingdom},
	title = {Piano Sustain-Pedal Detection Using Convolutional Neural Networks},
	pages = {241--245},
	booktitle = {Proceedings of {IEEE} International Conference on Acoustics, Speech and Signal Processing ({ICASSP})},
	author = {B. Liang and G. Fazekas and M. B. Sandler},
        address = {Brighton, United Kingdom},
	date = {2019-05},
        year = {2019},
}

@article{liang_measurement_2018,
	title = {Measurement, Recognition, and Visualization of Piano Pedaling Gestures and Techniques},
	volume = {66},
	pages = {448--456},
	number = {6},
	journal = {Journal of the Audio Engineering Society},
	shortjournal = {J. Audio Eng. Soc.},
	author = {B. Liang and G. Fazekas and M. B. Sandler},
	urldate = {2024-03-25},
	date = {2018-06-18},
        year = {2018},
        month = {June},
	langid = {english},
        address = {New York, United States},
}

@inproceedings{liang_piano_2017,
	location = {Copenhagen, Denmark},
	title = {Piano Pedaller: A Measurement System for Classiﬁcation and Visualisation of Piano Pedalling Techniques},
	pages = {325--329},
	booktitle = {Proceedings of The International Conference on New Interfaces for Musical Expression ({NIME})},
	author = {B. Liang and G. Fazekas and M. B. Sandler and A. {McPherson}},
	year = {2017},
	address = {Copenhagen, Denmark},
}

@inproceedings{liang_detection_2017,
	address = {New York, United States},
	title = {Detection of Piano Pedaling Techniques on the Sustain Pedal},
	booktitle = {Proceedings of Audio Engineering Society Convention 143},
	author = {B. Liang and G. Fazekas and M. B. Sandler},
	date = {2017-10},
	year = {2017}
}

@inproceedings{liang_transfer_2019,
	address = {Budapest, Hungary},
	title = {Transfer Learning for Piano Sustain-Pedal Detection},
	pages = {1--6},
	booktitle = {2019 International Joint Conference on Neural Networks ({IJCNN})},
	publisher = {{IEEE}},
	author = {B. Liang and G. Fazekas and M. B. Sandler},
	date = {2019-07},
	langid = {english},
	year = {2019}
}

@inproceedings{liang_piano_2018,
	title = {Piano legato-pedal onset detection based on a sympathetic resonance measure},
	booktitle = {Proceedings of the 26th European Signal Processing Conference ({EUSIPCO})},
	author = {B. Liang and G. Fazekas and M. B. Sandler},
	year = {2018},
	address = {Rome, Italy},
}

@inproceedings{yan_scoring_2024,
	address = {San Francisco, United States},
	title = {Scoring Time Intervals Using Non-Hierarchical Transformer for Automatic Piano Transcription},
	booktitle = {Proceedings of the International Society for Music Information Retrieval Conference ({ISMIR})},
	author = {Y. Yan and Z. Duan},
	year = {2024},
}

@inproceedings{yan_skipping_2021,
	title = {Skipping the Frame-Level: Event-Based Piano Transcription With Neural Semi-{CRFs}},
	volume = {34},
	pages = {20583--20595},
	booktitle = {Advances in Neural Information Processing Systems},
	author = {Y. Yan and F. Cwitkowitz and Z. Duan},
	year = {2021},
        address = {Virtual},
}

@inproceedings{Hawthorne2018,
  author    = {C. Hawthorne and E. Elsen and J. Song and A. Roberts and I. Simon and C. Raffel and J. Engel and S. Oore and D. Eck},
  title     = {Onsets and Frames: Dual-Objective Piano Transcription},
  booktitle = {Proceedings of the 19th International Society for Music Information Retrieval Conference (ISMIR)},
  year      = {2018},
  address  = {Paris, France}
}

@article{kong_high-resolution_2021,
	title = {High-Resolution Piano Transcription With Pedals by Regressing Onset and Offset Times},
	volume = {29},
	issn = {2329-9290},
	pages = {3707--3717},
	journal = {{IEEE}/{ACM} Transactions on Audio, Speech and Language Processing},
	shortjournal = {{IEEE}/{ACM} Trans. Audio, Speech and Lang. Proc.},
	author = {Q. Kong and B. Li and X. Song and Y. Wan and Y. Wang},
	urldate = {2024-04-08},
	date = {2021-10-26},
	year = {2021},
        month = {October},
        address = {New York, United States},
}

@book{banowetz_pianists_1985,
	address = {Bloomington, {IN}},
	title = {The Pianist's Guide to Pedaling},
	publisher = {Indiana University Press},
	author = {J. Banowetz},
	year = {1985},	
}

@inproceedings{raffel2014_mir_eval,
  author    = {C. Raffel and B. McFee and E. J. Humphrey and J. Salamon and O. Nieto and D. Liang and D. P. W. Ellis},
  title     = {{mir\_eval: A transparent implementation of common MIR metrics}},
  booktitle = {Proceedings of the International Society for Music Information Retrieval Conference (ISMIR)},
  year      = {2014},
  address   = {Taipei, Taiwan}
}

@article{poliner2007_melody_transcription,
  author  = {G. E. Poliner and D. P. W. Ellis and A. F. Ehmann and E. G{\'o}mez and S. Streich and B. Ong},
  title   = {Melody Transcription from Music Audio: Approaches and Evaluation},
  journal = {IEEE Transactions on Audio, Speech, and Language Processing},
  volume  = {15},
  number  = {4},
  pages   = {1247--1256},
  year    = {2007}
}

@article{salamon2014_melody_extraction,
  author  = {J. Salamon and E. G{\'o}mez and D. P. W. Ellis and G. Richard},
  title   = {Melody Extraction from Polyphonic Music Signals: Approaches, Applications, and Challenges},
  journal = {IEEE Signal Processing Magazine},
  volume  = {31},
  number  = {2},
  pages   = {118--134},
  month   = {March},
  year    = {2014}
}

@techreport{davies2009_beat_eval_tr,
  author      = {M. E. P. Davies and N. Degara and M. D. Plumbley},
  title       = {Evaluation Methods for Musical Audio Beat Tracking Algorithms},
  institution = {Centre for Digital Music, Queen Mary University of London},
  type        = {Technical Report},
  number      = {C4DM-TR-09-06},
  address     = {London, England},
  month       = {October},
  year        = {2009}
}

@inproceedings{bock2012_onset_online,
  author    = {S. B{\"o}ck and F. Krebs and M. Schedl},
  title     = {Evaluating the Online Capabilities of Onset Detection Methods},
  booktitle = {Proceedings of the International Society for Music Information Retrieval Conference (ISMIR)},
  year      = {2012},
  pages     = {49--54}
}

@article{ni2013_subjectivity_chord_eval,
  author  = {Y. Ni and M. McVicar and R. Santos-Rodriguez and T. De Bie},
  title   = {Understanding Effects of Subjectivity in Measuring Chord Estimation Accuracy},
  journal = {IEEE Transactions on Audio, Speech, and Language Processing},
  volume  = {21},
  number  = {12},
  pages   = {2607--2615},
  year    = {2013}
}

@inproceedings{pauwels2013_evaluating_chords,
  author    = {J. Pauwels and G. Peeters},
  title     = {Evaluating Automatically Estimated Chord Sequences},
  booktitle = {IEEE International Conference on Acoustics, Speech and Signal Processing (ICASSP)},
  pages     = {749--753},
  year      = {2013},
  publisher = {IEEE}
}

@inproceedings{turnbull2007_supervised_boundaries,
  author    = {D. Turnbull and G. Lanckriet and E. Pampalk and M. Goto},
  title     = {A Supervised Approach for Detecting Boundaries in Music Using Difference Features and Boosting},
  booktitle = {Proceedings of the International Society for Music Information Retrieval Conference (ISMIR)},
  pages     = {51--54},
  year      = {2007}
}

@article{levy2008_struct_seg_clustering,
  author  = {M. Levy and M. Sandler},
  title   = {Structural Segmentation of Musical Audio by Constrained Clustering},
  journal = {IEEE Transactions on Audio, Speech, and Language Processing},
  volume  = {16},
  number  = {2},
  pages   = {318--326},
  year    = {2008}
}

@inproceedings{lukashevich2008_segmentation_measures,
  author    = {H. M. Lukashevich},
  title     = {Towards Quantitative Measures of Evaluating Song Segmentation},
  booktitle = {Proceedings of the International Society for Music Information Retrieval Conference (ISMIR)},
  pages     = {375--380},
  year      = {2008}
}

@inproceedings{Fang25pedal,
  title={High-Resolution Sustain Pedal Depth Estimation from Piano Audio across Room Acoustics},
  author={K. Fang and H. Zhang and Z. Wang and I. Fujinaga},
  booktitle={Proceedings of the 26th International Society for Music Information Retrieval Conference (ISMIR)},
  year={2025},
  address={Daejeon, Korea},
  month={September},
  day={21--25}
}

@book{vonArx2014,
  author    = {V. A. von Arx},
  title     = {Piano Lessons with Claudio Arrau: A Guide to His Philosophy and Techniques},
  publisher = {Oxford University Press},
  year      = {2014},
  month     = {May},
  day       = {27},
  isbn      = {978-0199924349},
  pages     = {576},
  language  = {English},
  format    = {Paperback}
}

@inproceedings{asap-dataset,
  title={{ASAP}: a dataset of aligned scores and performances for piano transcription},
  author={F. Foscarin and A. McLeod and P. Rigaux and F. Jacquemard and M. Sakai},
  booktitle={International Society for Music Information Retrieval Conference {(ISMIR)}},
  year={2020},
  pages={534--541}
}

@techreport{emiya:inria-00544155,
  TITLE = {{MAPS - A piano database for multipitch estimation and automatic transcription of music}},
  AUTHOR = {V. Emiya and N. Bertin and B. David and R. Badeau},
  URL = {https://inria.hal.science/inria-00544155},
  TYPE = {Research Report},
  PAGES = {11},
  YEAR = {2010},
  MONTH = Jul,
  HAL_ID = {inria-00544155},
  HAL_VERSION = {v1},
}

\end{document}